\documentclass[amssymb,amsmath,aps,prb,showpacs]{revtex4-2}
\usepackage{array}
\usepackage{graphicx}
\usepackage{color}
\usepackage{multirow}
\usepackage{float}
\usepackage{amssymb}
\usepackage{amsmath}
\usepackage{makecell}
\usepackage[utf8]{inputenc}
\begin{document}

\title{On electrical transport and thermoelectric performance in half-Heusler phase ScNiSb}
\author{K. Ciesielski}
\author{Dariusz Kaczorowski}
\affiliation{Institute of Low Temperature and Structure Research, Polish Academy of Sciences, P. Nr 1410, 50-950 Wrocław, Poland,}

\begin{abstract}
Half-Heusler phases are among the most extensively studied thermoelectic materials. Bipolar thermal conductivity analysis performed for their sub-group based on rare-earth ($R$) metals, \textit{R}NiSb, indicated on high mobility ratio in favor of electrons. The suggestion found its experimental verification and led to significant improvement of thermoelectric properties in $n$-type doped series ScNiSb$_{1-x}$Te$_x$ \mbox{[\textit{Phys. Rev. Appl.} 15, 044047, (2021)]}. Recently, Kajikawa attempted alternative interpretation of transport properties in \mbox{ScNiSb}, where multi-parameter fit lead to mobility ratio in favor of holes [\textit{Int. J. Mod. Phys. B} 2250071 (2022)]. In this work we discuss the details of electrical properties and thermoelectric performance of ScNiSb in  context of structural disorder. The article considers also relevant assumptions regarding band degeneracy and scattering mechanisms for effective mass modeling in ScNiSb. Lastly, technical difficulties of the model proposed by Kajikawa are addressed. We believe, that the provided insight will be useful for understating  electrical transport in half-Heusler compounds, which can contribute to further improvement of their thermoelectric performance.

\end{abstract}


\maketitle

\section{Introduction}

Half-Heusler (HH) phases are intermetallics with general chemical composition \textit{MTX}, where \textit{M} is early transition metal or rare earth element, \textit{T} stands for late transition metal, and \textit{X} denotes $p$-electron element. They crystallize in cubic unit cell with space group $F\bar{4}3m$, no. 216. Compounds from this group are competitive\cite{zhu2015}$^-$\cite{quinn2021} with other extensively studied thermoelectrics\cite{dado2010}$^-$\cite{sadia2013}. HH materials are recognized especially for their excellent electrical properties, which stem from high band degeneracy ($N_v$), and high density-of-states effective masses ($m_{eff}$), accompanied by decent mobility of charge carriers ($\mu_H$)\cite{fu2014}. However, due to relatively simple unit cell, in  pristine (undoped) form they can exhibit rather high lattice thermal conductivity\cite{zhu2015}, which is undesirable from thermoelectric perspective.

The subgroup of HH compounds based on rare-earth elements has been attracting continuous attention of thermoelectric community for the last two decades\cite{sportouch1999}$^-$\cite{kong2012}.  Materials from this family are narrow band gap semiconductors ($E_g \lesssim 0.5$ eV)\cite{sportouch1999}$^,$\cite{ciesielski2020rnsib}. Their thermal conductivity in the proximity 300 K was shown to attain lower values (\textit{e.g.} 3.5 and 2.8 W/mK for ErPdSb\cite{gofryk2007} and YPtSb\cite{ouardi2011}, respective) than for classical HH thermoelectrics (\textit{e.g.} 6.3 and 8.0 W/mK for HfNiSn, and ZrNiSn, respectively)\cite{hohl1997}, when pristine, undoped compounds are considered. HH materials based on rare-earth elements were also shown to comprise extraordinary structural defects: atoms of late transition metal can be displaced from nominal 4$c$ Wyckoff position ($\frac{1}{4},\frac{1}{4},\frac{1}{4}$) to quarterly occupied 16$e$ ($x$,$x$,$x$) site, with parameter $x$ spanning from 0.256 to 0.262\cite{ciesielski2020hoptsb}$^,$\cite{synoradzki2019scnisb}. For the other HH compounds, different point defects are expected, \textit{e.g.} ZrNiSn can exhibit Ni interstitials in its unit cell\cite{xie2012}. Both rare-earth bearing as well as other HH materials show significant influence of disorder on their transport properties, see \textit{e.g.} Refs. \cite{gnida2021}, \cite{xie2014}.

Intrinsically low thermal conductivity and curious crystal structure motivated us to perform more extensive study of rare-earth containing HH materials with composition $R$NiSb, where $R$ is rare earth element\cite{ciesielski2020pra1}. We aimed at investigating the most important features driving electrical and thermal transport. As a result, we confirmed their narrow band gap, revealed that at low temperatures ionized impurity scattering might be important mechanism, and found that most of the studied materials are light-hole conductors. We also calculated the experimental charge carrier relaxation time, which was later on employed in further \textit{ab-initio} studies of $R$NiSb compounds\cite{satyan2022}$^,$\cite{Satyan2022APA}. It was also shown that the low values of observed lattice thermal conductivity in $R$NiSb compounds results from low sound velocity and intense scattering of phonons on point defects\cite{ciesielski2020pra1}. Last but not the least, we proposed a simple approach to analysis of bipolar thermal conductivity. Modeling of this phenomenon can provide information about mobility ratio ($b = \frac{\mu_{H(e)}}{\mu_{H(h)}}$, where $\mu_{H(e)}$ and $\mu_{H(h)}$ denote Hall mobility of electrons and holes, respectively). We predicted, that for $R$NiSb compounds, the  electrons should be characterized by higher mobility than holes. This feature suggests the likelihood of better thermoelectric performance after $n$-type doping.

In order to verify this conclusion, in separate work we performed $n$-type doping for one of the representatives of $R$NiSb family, $R$ = Sc. We shown that $\mu_H$ increase  around order of magnitude from $p$-type to $n$-type doped, when samples are synthesized with the same procedure (arc-melting)\cite{ciesielski2021rnisbPRA}. According to the predictions, we also achieved significant increase in thermoelectric performance in $n$-type ScNiSb.

Recently, Y. Kajikawa performed alternative analysis\cite{kajikawa2022} of our data for $p$-type ScNiSb from Ref. \cite{ciesielski2020pra1}. He claimed that it is possible to construct a model, in which holes have bigger mobility than electron in ScNiSb. For this means, Kajikawa introduced four scattering mechanisms for free holes (acoustic phonons, non-polar optical phonons, ionized impurities, and grain boundaries)  and added three more conductivity channels (free electrons, variable-range hopping and nearest neighbor hopping). Furthermore, validity of acoustic phonon scattering for modeling of thermoelectric properties was questioned. We appreciate interest in our work and effort invested in attempt of alternative interpretation. In this paper, we discuss in detail electronic transport of ScNiSb and evaluate the model proposed by Kajikawa\cite{kajikawa2022}. In the first section we will consider experimental validation of mobility ratio predicted for ScNiSb in context of structural disorder. The second section will assess model of transport properties proposed by Kajikawa and compare it with formerly reported analytical techniques.

\section{Results and discussion}
\subsection{Experimental validation of mobility ratio}

Modeling of bipolar thermal conductivity predicted the mobility ratio between electrons and holes in ScNiSb of $b$ = 7.9, see Ref. \cite{ciesielski2020pra1}. In our experimental validation, we observed increase of $\mu_H$ from 0.7 cm$^2$V$^{-1}$s$^{-1}$ for $p$-type ScNiSb to 6-17 cm$^2$V$^{-1}$s$^{-1}$ for $n$-type ScNiSb$_{1-x}$Te$_x$ \cite{ciesielski2021rnisbPRA}. Due to rather qualitative nature of the model and polycrystalline nature of the samples, the approximately order of magnitude increase in mobility was appraised as positive verification of our predictions. Kajikawa claimed, though, that our validation is not conclusive due to troubles with absolute values $\mu_H$\cite{kajikawa2022} -- in our first article on the topic\cite{ciesielski2020pra1}, we observed room temperature value of mobility in $p$-type ScNiSb of 24 cm$^2$V$^{-1}$s$^{-1}$, which is beyond the range of 0.7-17 cm$^2$V$^{-1}$s$^{-1}$ observed in Ref. \cite{ciesielski2021rnisbPRA}. We are grateful for raising this issue, as clarifying its context will possibly be helpful also for other readers.

Half-Heusler phases are known to be prone to crystallographic disor-der\cite{ciesielski2020hoptsb}$^-$\cite{xie2012}$^,$\cite{Kim2017}$^-$\cite{Pottgen2009}. Crystallographic imperfections were shown to have profound effects on mobility of charge carriers in this family of compounds\cite{xie2014}$^,$\cite{Xie2013}. Furthermore, the disorder can be synthesis-specific. In the case of ScNiSb, sample preparation by arc-melting, annealing and spark plasma sintering (SPS) lead to the crystallographic defects of Ni atomic displacement from their nominal 4$c$ Wyckoff position ($\frac{1}{4}$,$\frac{1}{4}$,$\frac{1}{4}$) to the quarterly occupied site 16$e$ with coordinates ($x$,$x$,$x$), x = 0.256\cite{ciesielski2020pra1}$^,$\cite{synoradzki2019scnisb}. No nickel vacancies were observed for this samples and the only detectable impurity phase was minuscule amount of Sc$_2$O$_3$\cite{ciesielski2020pra1}$^,$\cite{synoradzki2019scnisb}. In Ref. \cite{ciesielski2021rnisbPRA}, ScNiSb samples were prepared only by arc-melting and annealing; spark plasma sintering was here not included  due to our technical limitations. In these samples, we detected also small amounts of Ni-rich impurity phases and, consistently with this observation, preliminary refinement of lab-XRD data suggested also presence of minute amount of vacancies\cite{ciesielski2021rnisbPRA}. Vacancies on nickel site were previously reported by Harmening \textit{et. al} in single crystals of \mbox{ScNiSb}, which were crushed out of polycrystalline specimens prepared by arc-melting and annealing\cite{Pottgen2009}. The other example of synthesis-dependent disorder in half-Heusler family is ZrNiSn. The compound is known to exhibit Ni interstitials in polycrystalline form synthesized via conventional methods\cite{xie2012}$^,$\cite{Romaka2013}. Amount of the defects, though, can be substantially reduced via slow-cooling synthesis of single crystals\cite{Fu2020}.

Having said that, while the the most fundamental features of band structure in half-Heusler compounds should not drastically change from one series of samples to another synthesized by different methods, the absolute values of mobility can be affected by changing number of defects. Hence, the comparison of mobility between electrons and holes should be performed only, when $p$-type and $n$-type samples are prepared using the same synthesis route. Such comparison was performed in our work on system ScNiSb$_{1-x}$Te$_x$, which shown the expected higher mobility of electrons than holes\cite{ciesielski2021rnisbPRA}. Contrasting absolute value of mobility to $p$-type ScNiSb sample prepared by a different route, done by Kajikawa\cite{kajikawa2022} is not justified.

The second question regards the solution of the model for bipolar thermal conductivity, which provides, the same results for mobility ratio: $b$ and 1/$b$. Our choice of holes as carriers with smaller mobility was assessed by Kajikawa as speculative\cite{kajikawa2022}. In our article, however, we shown indication, which suggests that in the case of ScNiSb, holes are significantly more likely to exhibits smaller mobility\cite{ciesielski2020pra1}. The technique employs weighted mobility, $U = \mu_0\left(\frac{m_{eff}}{m_e}\right)^{3/2}$, where $\mu_0$ denotes intrinsic mobility, which can be approximated using parabolic band model, see Ref.~\cite{ciesielski2021rnisbPRA}. The other important parameter is band gap obtained from Goldsmid-Sharp equation: $E_g^* = 2eS_{max}T_{max}$, where $S_{max}$ and $T_{max}$ stand for the maximum Seebeck coefficient and the temperature at which it was obtained\cite{goldsmid1999}. The magnitude of $E_g^*$ should be compared with other values of the band gap, that can be considered more reliable. The difference between the two can occur for materials with largely different weighted mobility between electrons and holes. When the majority charge carriers have larger $U$, the Goldsmid-Sharp band gap will be overestimated with respect to real $E_g$\cite{snyder2015}$^,$\cite{felser2015}. It was shown, that behavior of half-Heusler phases \mbox{(Ti,~Zr,~Hf)NiSn} can be better described with weighted mobility ratio between electrons and holes equal to 5, see Ref. \cite{felser2015}.

In the case of $p$-type ScNiSb, we observed that Goldsmid-Sharp band gap (214~meV) is smaller than resistivity gap (383 meV). This observation strongly implies, that weighted mobility should be bigger for minority carriers (electrons in this case). The finding motivated us to indicate on mobility ratio $b$ in favor of electrons in ScNiSb\cite{ciesielski2020pra1}. Subsequent experimental validation in ScNiSb$_{1-x}$Te$_x$, confirmed predictions for $U$ changes with doping. The weighted mobility increased from the $p$-type sample to the most heavily $n$-type doped ScNiSb$_{0.85}$Te$_{0.15}$ 16-fold at temperature of optimal thermoelectric performance\cite{ciesielski2021rnisbPRA}.

The third and arguably most important perspective, which confirms validity of our modeling is thermoelectric performance in ScNiSb system. Both mobility and weighted mobility predictions are expected to guide experimentalists for the more promising direction of doping. The thermoelectric quality factor $B$ depends on fundamental parameters of material as $B \propto \frac{m_{eff}^{3/2}\mu_0}{\kappa_L}$,  where $\kappa_L$ stands for lattice thermal conductivity\cite{wang2013}. Values of $B$ determine maximum achievable figure of merit ($zT = \frac{S^2}{\rho\kappa}T$, where $\kappa$ stands for total thermal conductivity). Hence, increase in mobility or even more importantly in $U$, should be accompanied by rise of power factor ($PF =\frac{S^2}{\rho}$). The correlation between those parameters in ScNiSb$_{1-x}$Te$_x$ is indeed very strong. The values of $PF$ rise with concentration of $n$-type carriers in this system 17-fold, from 2.3 $\mu$Wcm$^{-1}$K$^{-2}$ $p$-type ScNiSb to 40 $\mu$Wcm$^{-1}$K$^{-2}$ for ScNiSb$_{0.85}$Te$_{0.15}$, both at \textit{ca.} 750~K in arc-melted samples\cite{ciesielski2020pra1}. This value of power factor is among the highest for experimentally known rare-earth bearing HH compounds, see \textit{e.g.} Refs. \cite{sportouch1999}-\cite{kong2012}, \cite{gofryk2008}-\cite{kawano2007APL}.

\subsection{Assessment of the modeling of transport properties}

Now that the doubts regarding direct experimental observations are dispelled, we can proceed to discussing details of the analysis preformed by in Ref. \cite{kajikawa2022} and questions therein regarding previous modeling of ScNiSb. In his work Kajikawa suggested alternative approach for description of transport properties in $p$-type ScNiSb with large number (18) of fit parameters\cite{kajikawa2022}. The formalism assumes that electronic transport in ScNiSb is a sum of four contributions: free holes, free electrons, nearest-neighbor hopping, and variable-range hopping. The scattering sources for free holes are ionized impurities, grain boundaries, acoustic phonons and non-polar optical phonons. The free coefficients are fitted simultaneously to Seebeck, resistivity, and Hall data. Using a model with such large number of parameters, most of which are refined to the same data set, is likely to result in several solutions similar in terms of fit quality.  Furthermore, not all values of fit parameters obtained by Kajikawa are reported in the article; the paper misses the value of $\Theta_{SVRH}$, which is described as a measure  of the asymmetricalness of electronic density of states. On top of that, the model requires 8 other material coefficients (\textit{e.g.} average and longitudinal sound velocity, deformation potential for acoustic phonons and non-polar optical phonons, etc.), that need to be obtained from experimental or theoretical literature on ScNiSb or approximated using knowledge for other related compounds.

On the contrary, in Ref. \cite{ciesielski2020pra1} we tried to minimize the number of fit parameters per curve; whenever possible we calculated  the transport parameters directly from the data, without fitting. The used experimental set included not only electrical transport, but also thermal conductivity and heat capacity; all data was experimental and was gathered for the very same specimen.

In short description of the procedure, the reduced chemical potential ($\eta$) in parabolic band model used in Ref. \cite{ciesielski2020pra1} is calculated as a single parameter in equation employing Seebeck data. With so-obtained $\eta$, we calculate directly the effective mass from the carrier concentration data. Then, the energy gap is obtained from two-parameter Arrhenius fit to resistivity curve. Later on, total thermal conductivity is subtracted by the polar electronic contribution  ($\kappa_{el-polar}$) calculated from Franz-Wiedemann law. The so-obtained data is used for Callaway model with four scattering coefficients. In the above analysis, we also use Debye temperature ($\Theta_D$), which is obtained from the heat capacity data. Value of $\Theta_D$ is used for calculation of sound velocity. As soon as lattice thermal conductivity is obtained, we proceed to calculating the bipolar thermal conductivity as $\kappa_{el-bipolar}$ = $\kappa$ - $\kappa_L$ - $\kappa_{el-polar}$. Most importantly, the result of the subtraction is used to perform fit in which $b$, the mobility ratio, is the only refined parameter. Simplicity of the approach, smaller amount of simultaneously used free coefficients and fully experimental data gathered for the same specimen (let alone the same compound) are expected to reduce the margin of error. The readers are referred to Ref. \cite{ciesielski2020pra1} for more detailed description of the analytic procedure.

Further on, we will discuss several more specific features of the used models. Kajikawa noted that for the two $p$-type ScNiSb samples our effective masses at 300 K differed significantly. The values pointed out were 1.52$m_e$\cite{ciesielski2020pra1} and 0.10$m_e$\cite{ciesielski2021rnisbPRA} for the ScNiSb specimens synthesized with and without spark plasma sintering, respectively. The former value is close to the expectations from ab-initio calculations (1.19$m_e$)\cite{winiarski2018}. Here, we would like to remind, that the latter value of $m_{eff}$ (0.10$m_e$) was assessed in our article as an artifact stemming from proximity of Seebeck coefficient reaching the value of zero near room temperature\cite{ciesielski2021rnisbPRA}. The transitions between $n$-type and $p$-type conductivity in semiconductors with low carrier concentration are frequently observed phenomena. When we approximated the effective mass at 380 K for $p$-type ScNiSb synthesized without SPS, \textit{i.e.} only slightly further from the point plausible transition in thermopower, the effective mass was clearly reaching closer to the expectations, 0.49$m_e$ with parabolic band model, and 0.68$m_e$ for $m_{eff}$ calculated from the equation for weighted mobility\cite{ciesielski2021rnisbPRA}.

Next, we would like to refer to the issue of band degeneracy in ScNiSb. In Ref.~\cite{ciesielski2020pra1} the number of valleys ($N_v$) for valence band of ScNiSb  was assumed as 6, taking into account both orbital and spin degeneracy. We agree with Kajikawa, that assuming only orbital degeneracy ($N_v$ = 3) is sufficient\cite{kajikawa2022}. Taking into account this approach, the \textit{ab-initio} \textit{p}-type effective mass of ScNiSb would be equal to $m_{eff} = N_v^{2/3}m_b = 0.75m_e$, which is, as expected, still in range between two experimentally obtained values of $m_{eff}$ for p-type ScNiSb\cite{ciesielski2020pra1}$^,$\cite{ciesielski2021rnisbPRA}. During preparation of this work we also noticed a typo in Ref. \cite{ciesielski2021rnisbPRA} in the related topic. Value of $N_v$ for valence band of ScNiSb was denoted as 2, where the correct value is 3. This parameter was not employed in any stages of the modeling of transport properties, hence it does not influence the core findings of the article.

Further on, it is necessary to discuss with special attention the charge carrier scattering mechanisms in ScNiSb. Kajikawa criticized, that the   parabolic band analysis in our articles was performed with assumption of acoustic phonons as dominant scattering sources\cite{kajikawa2022}. The \textit{ab-initio} derived deformation potential, which described strength of electron-phonon coupling was reported as equal to 0.37~eV for valence band of ScNiSb\cite{zhou2018}. This value is  lower than the expected range of 5-35 eV for typical semiconductors\cite{Xie2013}. Based on this feature, Kajikawa claimed that the interaction between electrons and acoustic phonons will not be the dominant scattering mechanism in ScNiSb\cite{kajikawa2022}. It was also reminded that ScNiSb exhibits in range 2-200 K rise in mobility with increasing temperature \cite{ciesielski2020pra1}, which is suggestive of grain boundary scattering or ionized impurity scattering (IIS). The latter mechanism appeared more likely even slightly above 200 K, due to observations of rising mobility with temperature up to \textit{ca.} 320 K in single crystals of related HH phase, ZrNiSn\cite{ren2020}.

Half-Heusler phases usually show maximum thermoelectric performance above 600 K, see \textit{e.g.} Refs. \cite{fu2014}, \cite{Li2013}, \cite{ciesielski2020rnsib}. At significantly elevated temperature range, $T>$~500~K, the combination of scattering on acoustic phonons and point defects is expected to dominate over IIS or effect of grain boundaries, which behavior was frequently observed in the literature on half-Heusler phases and other thermoelectric materials, see \textit{e.g.} Refs. \cite{xie2014}, \cite{fu2013}-\cite{snyder2020}.  Based on these reasons, we applied the scattering mechanisms in our calculations, that describe  acoustic phonon and point defect scattering and are most likely to be appropriate for temperatures of maximum performance on ScNiSb. Furthermore, it is worthy noting, that ScNiSb shown signs of transition to scattering on acoustic phonons and point defects even in the already present data Hall data ($T \leq $~300 K), as its mobility starts to decrease with rising $T$ above \textit{ca.} 225 K.

We also underline, that acoustic phonon scattering  is described in parabolic band modelling by the same scattering coefficient as interaction of charge carriers with point-defects. Hence, even if coupling between electrons or holes and acoustic phonons was small in ScNiSb, point defects are likely to be highly relevant as dominant scattering source for a HH compound due to crystallographic disorder. Dominance of point-defect scattering at elevated temperatures for half-Heusler phases was shown previously\cite{Xie2013}, including samples, which show IIS or grain boundary influence at low temperatures\cite{xie2014}. All the above considerations justify our approach to parabolic band modeling. As we guess, they were also at least among the reasons for which Kajikawa despite extensive critique of acoustic phonon scattering, used this scattering mechanism as dominant for calculations of Seebeck coefficient and Hall factor for free electrons in his own model\cite{kajikawa2022}.

\section{Summary}

In this article, we discuss the analysis of electronic transport properties and thermoelectric performance in ScNiSb. Special focus was paid to experimental validation of the mobility ratio between electrons and holes. The most important findings are following:

1) Contrasting mobility of samples obtained with different synthesis routes, as presented in Ref. \cite{kajikawa2022}, is unjustified due to the synthesis-dependent crystallographic disorder. Comparison withing the series of specimens obtained with the same synthesis methods proves mobility ratio favoring electrons in ScNiSb\cite{ciesielski2021rnisbPRA}.

2) Choice between two solutions of bipolar thermal conductivity ratio ($b$ and 1/$b$) was not unsubstantiated, as suggested by Kajikawa\cite{kajikawa2022}. Rather than that, it was based on indication from weighted mobility ratio in ScNiSb. The subsequent study confirmed the modeling; 16-fold increase in weighted mobility was observed from $p$-type sample to the most heavily $n$-type doped ScNiSb.

3) Thermoelectric performance in ScNiSb system serves as additional verification of analysis based on bipolar thermal conductivity. According to expectations, power factor for ScNiSb increased significantly (17-fold) after $n$-type doping, achieving value of 40 $\mu$Wcm$^{-1}$K$^{-2}$ at 740 K for ScNiSb$_{0.85}$Te$_{0.15}$, which is among the highest for rare-earth half-Heusler phases.

4) Model proposed in Ref. \cite{kajikawa2022} was shown to suffer from several technical difficulties, regarding primarily large numbers of simultaneously fitted parameters and material coefficients obtained from different sources (experiments or calculations for ScNiSb, and approximations from other compounds). For parabolic band modeling performed previously\cite{ciesielski2020pra1}, the doubts regarding differing values of effective masses for ScNiSb synthesized by distinct methods were clarified performing calculations further away in temperature scale from $n$-type to $p$-type transition in Seebeck coefficient. Lastly, assumption of acoustic phonon and point defects as dominant scattering sources in ScNiSb was justified as relevant for high temperature considerations.

Overall, the article shows, that despite thermoelectric materials can be considered as complex, it is usually beneficial to approach analysis of transport properties with the simplest available formalism. The above is especially important, if outcome of the modeling is aimed at guiding optimization of thermoelectric performance.

\section*{Acknowledgements}

This work was supported by the National Science Centre (Poland) under research grant 2021/40/Q/ST5/00066. For the purpose of Open Access, the author has applied a CC-BY public copyright licence to any Author Accepted Manuscript (AAM) version arising from this submission

\end{document}